\pdfoutput=1
\documentclass[aps,prx,floatfix,nofootinbib,superscriptaddress,twocolumn,10pt]{revtex4-2}
\usepackage{graphicx,graphics}
\usepackage{dcolumn}
\usepackage{newtxtext}
\usepackage{pifont}
\usepackage{amsmath,amssymb,amsfonts}
\usepackage{latexsym,verbatim}
\usepackage{bm,dsfont}
\usepackage[caption=false]{subfig}
\usepackage{xcolor}
\usepackage{multirow}
\usepackage{overpic}
\usepackage[colorlinks=true,linkcolor=blue!60!black,citecolor=blue!60!black, urlcolor=blue!60!black]{hyperref}
\usepackage{array}
\usepackage{nicefrac}
\usepackage{braket}
\usepackage{tikz}
\usepackage{quantikz}
\usepackage[most]{tcolorbox}
\usepackage{float}
\usepackage[normalem]{ulem}

\DeclareMathOperator{\diag}{diag}
\newcommand{\cmark}{\ding{51}}
\newcommand{\xmark}{\ding{55}}
\usepackage{amsthm}
\theoremstyle{definition}

\begin{document}

\title{Cavity-mediated probabilistic magic $T$-gate injection}

\author{Sofia Cocciaretto}
\email{sofia.cocciaretto@sns.it}
\affiliation{Scuola Normale Superiore, I-56127 Pisa, Italy}

\author{Roberto Menta}
\email{roberto.menta@sns.it}
\affiliation{NEST, Scuola Normale Superiore, I-56126 Pisa, Italy}

\author{Vittorio Giovannetti}
\email{vittorio.giovannetti@sns.it}
\affiliation{NEST, Scuola Normale Superiore, I-56126 Pisa, Italy}

\begin{abstract}
Non-Clifford gates are a necessary resource for universal quantum computation,
yet their fault-tolerant implementation typically relies on magic-state
distillation, which incurs significant overhead in qubit count and circuit depth. In this work, we propose a probabilistic cavity-based magic-state injection protocol.
Our scheme exploits controlled atom-cavity interactions and conditional
measurements to probabilistically prepare an effective magic state encoded in the first two level Fock subspace of a single cavity mode, achieving a success
probability of 0.74 per attempt, independent of the target
magic phase.
The cavity-encoded magic state is subsequently injected into a computational atom
via a teleportation-based protocol mediated by dressed-state transitions,
requiring only Clifford operations and a single auxiliary atom for readout.
We show that all required operations -- state preparation, two-qubit exchange
gates, and projective measurement -- can be implemented with experimentally
available techniques in Rydberg atom-cavity platforms.
We further discuss how the scheme can in principle be adapted to
operate at the logical level, where collective Rydberg interactions and
optical nonlinearities provide a route toward cavity-mediated $T$-gate
injection directly into code-encoded qubits.
\end{abstract}

\maketitle

\emph{\bfseries Introduction.}
Fault-tolerant quantum computation requires the ability to implement a universal
set of logical gates while suppressing physical errors below a threshold. In most
quantum error-correcting codes, including the surface
code~\cite{Kitaev2002, Raussendorf2007}, all Clifford gates $\{H,S,\mathrm{CNOT}\}$ admit
fault-tolerant implementations through transversal or locality-preserving
operations~\cite{Gottesman1997}. In contrast, non-Clifford gates -- most notably the $T$ gate  -- cannot be realized in this manner. This limitation is formalized by the
Eastin-Knill theorem~\cite{PhysRevLett.102.110502}, which forbids universal sets
of transversal gates, and by the Bravyi-K\"onig
theorem~\cite{PhysRevLett.110.170503}, which constrains the logical gates
achievable by locality-preserving operations in topological codes.
A standard strategy to circumvent these no-go results is \emph{magic state
injection}~\cite{GottesmanChuang1999, Zhou2000}. In this paradigm, a non-stabilizer
auxiliary state, a \emph{magic state}, is prepared and consumed via Clifford
operations and measurements to effectively teleport a non-Clifford gate onto
the data qubits. 
In practice, raw magic states are typically noisy and must be purified through
\emph{magic state distillation}~\cite{bravyi2005universal, Souza2011, SalesRodriguez2025}, a process that consumes many copies of imperfect states
and requires large Clifford circuits. As a result, magic state distillation
dominates the resource overhead of fault-tolerant quantum computing, both in
terms of qubit count and circuit depth~\cite{Fowler2012,Litinski2019}, and is widely recognized as a major
bottleneck for scalable quantum computation.

Recent years have seen substantial progress toward reducing or eliminating this
overhead~\cite{Campbell2017,Gupta2024,Wills2025}. One prominent direction is \emph{magic
state cultivation}~\cite{gidney2024magic,rosenfeld2025magic}, where structured
dynamical protocols directly prepare high-quality non-stabilizer states.
Related efforts include distillation-free or low-overhead constructions based on
tailored code structures~\cite{Scruby2025,fazio2025low} and
measurement-free protocols~\cite{Sascha2025}. On the experimental side, logical
magic states with fidelities beyond the distillation threshold have been
demonstrated on superconducting
platforms~\cite{PhysRevLett.131.210603}.
In this work, we contribute to this landscape by proposing a
\emph{cavity-based magic-state injection protocol}.
Our scheme prepares a physical magic state in the Fock subspace of a single cavity mode through a controlled atom-cavity interaction and a post-selection measurement.
By restricting the encoding to the $\ket{n=0,1}$ Fock manifold,
we achieve a success probability of approximately $74\%$, independent of the
target magic phase. The prepared cavity-encoded magic state is then injected
into a computational atom via a teleportation-based protocol requiring only
Clifford operations and a single auxiliary atom for readout. The protocol is naturally suited to Rydberg atom-cavity platforms~\cite{PRXQuantum.3.030314, 4c33-b1dv, de2026realization}, where
coherent-state preparation, adiabatic atomic transport, cavity-mediated
interactions, and high-fidelity projective measurements are experimentally
available. A single cavity and a single auxiliary atom suffice to inject
non-Clifford gates sequentially across an entire processor, offering a
resource-efficient alternative to large-scale distillation
factories. {We further discuss how the same principles extend, via
collective Rydberg-Rydberg interactions and optical nonlinearities, to
cavity-mediated $T$-gate injection at the logical level.}

We start by reviewing the dressed-state structure of atom-cavity systems. We then
present the cavity-mediated injection protocol and analyze its success
probability. We conclude discussing architectural considerations and future
directions. Appendix~\ref{app:simulation} collects all simulation parameters and a verification plot.

\begin{figure}[t!]
    \centering
    \includegraphics[width=\columnwidth]{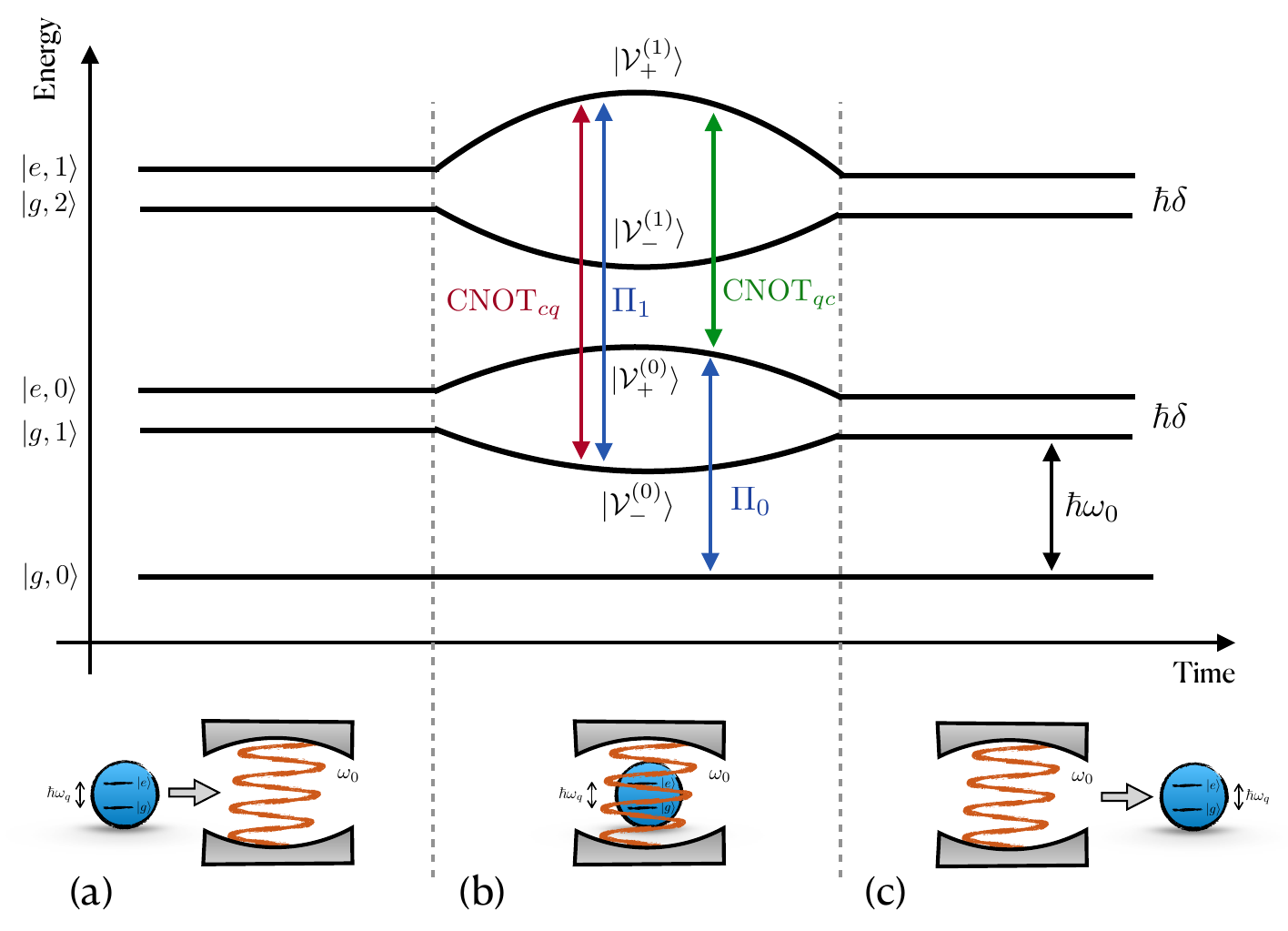}
    \caption{Instantaneous energy levels of the dressed states for $\delta>0$.
    The three time windows correspond to (a) atom outside the cavity
    ($\Omega\to 0$), (b) atom inside the cavity ($\Omega\sim\Omega_0$), and
    (c) atom outside again. The arrows indicate the resonant transitions used in
    the protocol.}
    \label{fig:dressed}
\end{figure}

\emph{\bfseries Quantum operations via driven JC model.}
Consider a two-level atom (qubit) with ground and excited states, $\ket{g}$ and $\ket{e}$ respectively,
separated by a transition frequency $\omega_q$,  coupled to the single mode of a cavity QED system. The dynamics  is described by the 
Jaynes-Cummings (JC) Hamiltonian~\cite{JaynesCummings1963,Brune1996}:
\begin{align}
    H :=& \dfrac{\hbar\omega_0}{2}\left(a^\dagger a + aa^\dagger\right)
    + \dfrac{\hbar \omega_q}{2}\left(\ket{e}\bra{e} - \ket{g}\bra{g}\right) \label{ham}\\ &+ \hbar \Omega(x)\left(\ket{e}\bra{g} a + \ket{g}\bra{e}a^\dagger \right) , \nonumber 
\end{align}
where $\omega_0$ is the frequency of the cavity mode, $a$ ($a^\dag$) is the  annihilation 
(creation) operator of the cavity mode, 
and   $\Omega(x)$ is the position-dependent vacuum Rabi frequency. Here, $x$ denotes the transverse displacement of the atom from the cavity axis. While the precise spatial dependence of $\Omega(x)$ is geometry dependent, for definiteness we model it as a Gaussian profile, $\Omega(x) = \Omega_0\,e^{-(x/D)^2}$, where $D$ characterizes the transverse extent of the cavity mode. This captures the exponential suppression of the atom-field coupling, and hence of the vacuum Rabi splitting, as the atom is displaced away from the cavity axis. In the following, we assume that 
the atom traverses the cavity, as illustrated in the bottom panel of Fig.~\ref{fig:dressed}, experiencing different values of $\Omega(x)$ along its trajectory. For a constant velocity
$v$ and sufficiently slow motion to justify the adiabatic approximation~\cite{HarocheRaimond2006,Walther2006}, 
the dynamics is described by an effective time-dependent Hamiltonian 
 $H_{\rm{eff}}(t)$ obtained from~(\ref{ham}) by
replacing  $\Omega(x)$ with 
\begin{equation}
    \label{equation:rabi_pulse}
    \Omega_{\rm{eff}}(t) \simeq \Omega(vt) =\Omega_0\,e^{-(t/\tau)^2},
\end{equation}
where $\tau:=D/v$ sets the characteristic interaction time.
While the adiabatic regime requires $\tau$ to be large compared to the inverse Rabi splitting, optimal control techniques~\cite{Glaser2015,Koch2022} make it possible to engineer the atomic injection protocol into the cavity so as to realize the same dressed-state operations over substantially shorter interaction times, reducing decoherence while maintaining high gate fidelity.
When the atom is far from the cavity, such that the interaction is negligible, the eigenstates of the system are the bare states $\ket{g,n} := \ket{g} \otimes \ket{n}$ and $\ket{e,n} := \ket{e} \otimes \ket{n}$, where $\ket{n}$ denotes the $n$-photon  Fock state of  the cavity mode.
As the atom enters the cavity, $\ket{g,0}$ remains an eigenstate, whereas all other bare states hybridize pairwise into the dressed states $\ket{\mathcal{V}_{\pm}^{(n)}(t)}$~\cite{Giovannetti2000}:
\begin{equation}
\label{eq:dressed}
\resizebox{\columnwidth}{!}{$
\ket{\mathcal{V}_{\pm}^{(n)}(t)}
=
\frac{
\Bigl(\delta/2 \pm \sqrt{(\delta/2)^2+\Omega_{\rm{eff}}^2(t)(n+1)}\Bigr)\ket{e,n}
+\Omega_{\rm{eff}}(t)\sqrt{n+1}\ket{g,n+1}
}{
\sqrt{
\delta^2/2+2\Omega_{\rm{eff}}^2(t)(n+1)
\pm \delta\sqrt{(\delta/2)^2+\Omega_{\rm{eff}}^2(t)(n+1)}
}
}
,
$}
\end{equation}
where $\delta:=\omega_q-\omega_0$ is the cavity-atom detuning. The corresponding
instantaneous eigenenergies are 
\begin{equation}
    E_{\pm}^{(n)}(t) = \hbar\omega_0(n+1)
    \pm\hbar\sqrt{(\delta/2)^2+\Omega_{\rm{eff}}^2(t)(n+1)}.
\end{equation}
As shown in the top panel of Fig.~\ref{fig:dressed}, the dressed-state spectrum is anharmonic, reflecting the $\sqrt{n+1}$ nonlinearity of the JC model. This anharmonicity enables selective addressing of individual transitions and thereby the implementation of nontrivial atom-cavity
quantum logical operations~\cite{PhysRevLett.74.4087,Giovannetti2000}.
In particular, two-qubit gates can be realized by encoding one qubit in a selected cavity Fock subspace, e.g., $\{|n\!=\!0\rangle, |n\!=\!1\rangle\}$, while the atom provides the second qubit~\cite{PhysRevLett.74.4087}. After the atom is adiabatically transported into the cavity, for instance by means of an optical conveyor belt, logical operations are implemented by applying resonant $\pi$-pulses
generated by a classical driving field:
\begin{equation}
    \label{equation:CNOT_pulse}
    H_{\rm field}(t)=-\hbar\Theta_s \cos(\omega_f t+\phi_f)
    e^{-(t/\tau_s)^2}
    \bigl(\ket{e}\bra{g}+\ket{g}\bra{e}\bigr),
\end{equation}
tuned to the splitting $\omega_f = E_{\pm}^{(n)}-E_{\pm}^{(m)}$ of the target
dressed-state pair (Fig.~\ref{fig:dressed}). Following the pulse, the atom is adiabatically extracted from the cavity, mapping the dressed-state dynamics back onto the bare-state basis.
As a representative example, a CNOT gate with the cavity-qubit $\{|n\!=\!0\rangle, |n\!=\!1\rangle\}$
as control and the
atom as target, $\mathrm{CNOT}^{(1)}_{c\to a}$,  is realized by resonantly driving the transition
$\ket{g,1}\leftrightarrow\ket{e,1}$ corresponding to $\ket{\mathcal{V}_{-}^{(0)}}\leftrightarrow\ket{\mathcal{V}_{+}^{(1)}}$
at frequency $\omega_f=E_+^{(1)}-E_-^{(0)}$. 
The anharmonicity of the JC ladder ensures spectral selectivity, allowing the targeted transition to be addressed while suppressing unwanted excitations.

\begin{figure}[t!]
    \centering
    \begin{overpic}[width=\columnwidth]{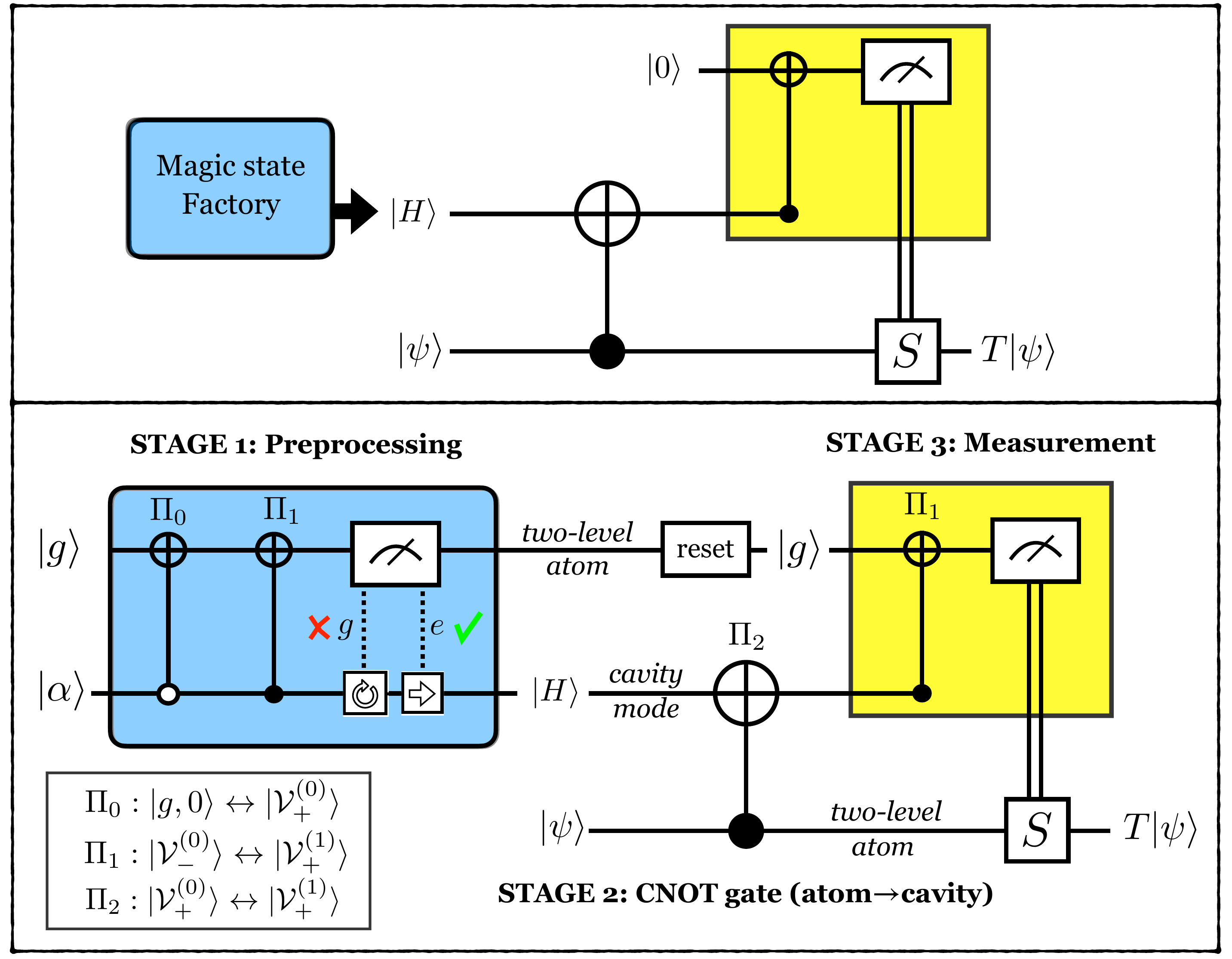}
    \end{overpic}
    \caption{Top panel: Magic-state injection circuit for the $T$ gate. The measurement of the auxiliary qubit prepared in the magic state $\ket{H}$ is implemented indirectly (yellow box) via a second auxiliary qubit initialized in $\ket{0}$. Bottom panel: cavity-mediated $T$ gate injection. The cavity (center) replaces the distillation factory: steps 1-3, through an auxiliary atom (top), prepare the magic state probabilistically, after which two CNOT gates and a single auxiliary measurement complete the injection  on the physical computational atom (bottom). The auxiliary atom is reinitialized and reused for readout.}
    \label{fig:T-injection}
\end{figure}

\emph{\bfseries Cavity-mediated $T$-gate injection protocol.}
The standard magic-state injection protocol for the 
$T:= \diag\left(1, e^{i\pi/4}\right)$ gate,
as applied for instance to the surface code~\cite{injection_surfacecode}, is described in 
the top panel of Fig.~\ref{fig:T-injection}. The state $\ket{\psi}=\alpha\ket{0}+\beta\ket{1}$ denotes the code qubit on which the $T$ gate is to be applied, while  \begin{eqnarray}\ket{H}:=    T\ket{+} = ({\ket{0}+e^{i\pi/4}\ket{1}})/{\sqrt{2}},\end{eqnarray}
is the corresponding magic state prepared on an auxiliary qubit by an external magic-factory
(hereafter $\ket{+}=(\ket{0} +\ket{1} )/\sqrt{2}$). Applying a CNOT gate to the joint state  $\ket{H}\ket{\psi}$ produces an equally weighted superposition 
$(\ket{0}T\ket{\psi}
+
e^{i\pi/4}  \ket{1} S^{\dag} T\ket{\psi})/\sqrt{2}$,
where $S^\dag$ is the adjoint of the (Clifford) phase gate 
$S:= \diag\left(1,i\right)$.
A projective measurement of the auxiliary qubit then yields either $T\ket{\psi}$ directly, upon 
obtaining the outcome $\ket{0}$, or $S^{\dag} T\ket{\psi}$, upon obtaining $\ket{1}$. In the latter case, the target state is recovered by applying the Clifford correction $S$.
Therefore measuring the auxiliary qubit in $\ket{0}$ yields $T\ket{\psi}$ directly, while measuring
$\ket{1}$ leads to the same result upon the application of $S$.
In the implementation proposed below, the projective measurement is performed indirectly via teleportation, by first transferring the state of the auxiliary qubit to a second ancillary qubit which is subsequently 
measured (yellow box in Fig.~\ref{fig:T-injection}).
More generally,  the protocol can be used to inject the phase gate
 $M_k:=\diag(1,e^{i\pi/(2k)})$ by preparing the auxiliary qubit in the corresponding magic
 state $\ket{M_k}:=M_k\ket{+}$ and applying the conditional correction
  $R_z(\pi/k)\propto\diag(1,e^{i\pi/k})$ following the measurement outcome
\cite{bravyi2005universal,reichardt2006quantum,
heinrich2019robustness}. While for $k=2$ the correction is the Clifford gate $S$, for $k>2$ it becomes non-Clifford and must itself be implemented through an additional round of magic-state injection or accounted for within a generalized Pauli frame~\cite{bravyi2005universal}.

We now introduce our cavity-mediated protocol for implementing $T$  gates via magic-state injection on a qubit 
represented by a two-level atom. In our scheme the magic state is encoded into a cavity mode that acts as the primary auxiliary qubit. The cavity thus plays a dual role, serving both as the magic-state resource and as the interaction region where the two-qubit gates required by the injection protocol are performed.
As shown in the bottom panel of Fig.~\ref{fig:T-injection}, the protocol comprises three stages. In a preprocessing step, the magic state $\ket{H}$ in a superposition of $\{|n\!=\!0\rangle, |n\!=\!1\rangle\}$ cavity states is produced by post-selecting a suitably chosen coherent state $\ket{\alpha}$ using an auxiliary atom. The information-carrying atom is then coupled to the cavity mode through a CNOT operation, implementing the coherent part of the magic-state injection protocol. Finally, the cavity is measured with a second auxiliary atom, completing the injection procedure.

\emph{Stage 1: Preprocessing.} To produce   the magic  resource state  $\ket{H}$ in the cavity, 
an auxiliary atom is prepared in $\ket{g}$ and inserted into the cavity initialized into a coherent state $\ket{\alpha}$ whose complex amplitude is defined in the following. The 
initial joint state can thus be expressed as
\begin{equation}
    \label{equation:psi_in}
    \ket{\Psi_{\rm in}} = \ket{g}\otimes \ket{\alpha}=  \ket{g}\otimes
    \Bigl(\beta_0\ket{0}+\beta_1\ket{1}+\sum_{n\geq 2}\beta_n\ket{n}\Bigr),
\end{equation}
with  $\beta_n := \frac{\alpha^n}{\sqrt{n!}}\,e^{-|\alpha|^2/2}$.
Two resonant $\pi$-pulses are then applied:
\begin{align}
    &\Pi_0:\quad\ket{g,0}\leftrightarrow\ket{\mathcal{V}_+^{(0)}}\;(\ket{e,0}),\\
    &\Pi_1:\quad\ket{\mathcal{V}_-^{(0)}}\;(\ket{g,1})
             \leftrightarrow\ket{\mathcal{V}_+^{(1)}}\;(\ket{e,1}).
\end{align}
Notice that $\Pi_1$ is precisely the $\mathrm{CNOT}^{(1)}_{c\to a}$ discussed in the previous paragraph; $\Pi_0$ instead corresponds to a CNOT gate where again the cavity qubit 
is the controller and the
atom the target, $\mathrm{CNOT}^{(0)}_{c\to a}$,  but with  the trigger state 
being the $|n\!=\!0\rangle$ Fock state instead of  $|n\!=\!1\rangle$.
As a consequence $\ket{\Psi_{\rm in}}$ gets transformed into 
\begin{align}
    \label{eq:msp-2}
    \ket{\Psi_{\rm out}}
    &= \Pi_1\Pi_0\ket{\Psi_{\rm in}} \nonumber\\
    &= \ket{e}\otimes(\beta_0\ket{0}+\beta_1\ket{1})
      +\ket{g}\otimes\sum_{n\geq 2}\beta_n\ket{n}.
\end{align}
We now perform a projective measurement on the auxiliary qubit: outcome $\ket{e}$ is accepted (\cmark); outcome $\ket{g}$
triggers a restart (\xmark). The success probability is
\begin{equation}
    P_e = |\beta_0|^2+|\beta_1|^2,
\end{equation}
and the accepted cavity state is $(\beta_0\ket{0}+\beta_1\ket{1})/\sqrt{P_e}$.
To match $\ket{H}$, the Fock amplitudes must satisfy the condition 
\begin{equation}
    \alpha= \frac{\beta_1}{\beta_0}
    \stackrel{!}{=}{e^{i\pi/4}}, 
\end{equation}
leading to 
\begin{equation}
    \label{equation:beta_vals}
    \beta_0 = \frac{1}{\sqrt{e}},\quad
    \beta_1 = \frac{e^{i\pi/4}}{\sqrt{e}},\quad
    P_e = \frac{2}{e}\approx 0.74.
\end{equation}
Each preparation attempt succeeds with probability $P_e=2/e$; upon failure, the cavity is reinitialized in the coherent state 
$\ket{\alpha=e^{i\pi/4}}$ and the preparation sequence is repeated. Since successive attempts are independent, the probability of success within at most $k$ attempts is
\begin{equation}
P_{\leq k}=1-\left(1-\frac{2}{e}\right)^k,
\end{equation}
yielding $P_{\leq1}\approx 74\%$, $P_{\leq2}\approx 93\%$, and $P_{\leq3}\approx 98\%$. The mean number of attempts is only $1/P_e\approx1.36$.

\begin{figure}[t]
    \centering
    \includegraphics[width=1\columnwidth]{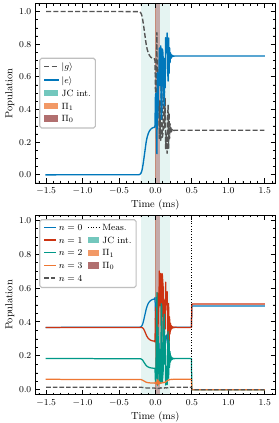}
    \caption{Simulation of the preprocessing stage of the protocol. The parameters defining  the effective (time-dependent) JC 
    Hamiltonian $H_{\rm{eff}}(t)$ are 
     $\Omega_0 = 420\,\mathrm{kHz}$, 
    $\tau = 100\,\mu\mathrm{s}$, and 
    $\delta =  0.18\,\Omega_0$. 
    The pulses $\Pi_0$ and $\Pi_1$ are activated with a relative delay of
    $30$ $\mu s$ setting the parameters of their associated $H_{\rm field}(t)$ as
    $(\Theta_0, \tau_0) = 
    (129.33\,\mathrm{kHz}, 3.8\,\mu\mathrm{s})$ and 
    $(\Theta_1, \tau_1) = (141.5\,\mathrm{kHz}, 19\,\mu\mathrm{s})$ 
    respectively. Top panel:  populations of the auxiliary atom after the coupling with cavity mode. The probability of getting the outcome $\ket{e}$
    is $\simeq 0.74$. Bottom panel: cavity Fock-level populations throughout the preprocessing stage. 
    Post-selecting on the $\ket{e}$ outcome, levels $\ket{0}$ and $\ket{1}$ carry equal population $\simeq 0.5$ and all $n\geq 2$ components vanish. 
}
\label{fig:cavity_population}
\end{figure}

\begin{figure}[t!]
    \centering
    \includegraphics[width=0.9\columnwidth]{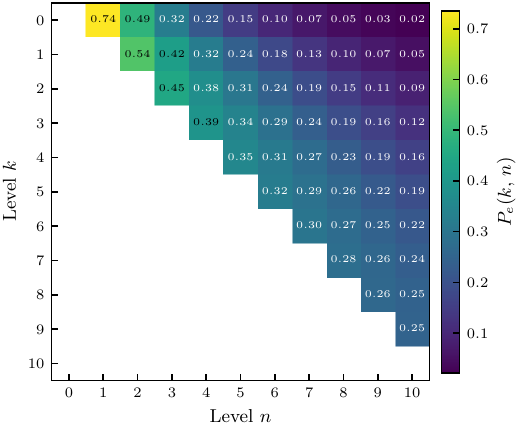}
    \caption{Success probability $P_e(k,n)$ when encoding state $\ket{H}$ in Fock levels
    $\ket{k}$ and $\ket{n}$ for all pairs $(k,n)$ among the
    first 10 cavity levels. The $\ket{n=0,1}$ encoding (top-left corner) achieves
    the global maximum $P_e=2/e\approx 0.74$.
   }
    \label{fig:prob_succ}
\end{figure}

In Fig.~\ref{fig:cavity_population} we report a numerical simulation of the process using parameters matching the CNOT implementation of Ref.~\cite{Giovannetti2000} (see Appendix~\ref{app:simulation} for details). 
An alternative approach to encode $\ket{H}$ in the cavity mode is to use $\ket{k},\ket{n}$ ($k<n$) levels instead of
$\ket{0}, \ket{1}$. With this choice the condition
$|\beta_k|=|\beta_n|$ forces us to have  $|\alpha|^2=\Lambda^{1/(n-k)}$ with
$\Lambda=n!/k!$, so that the success probability reduces to
\begin{equation}
    \label{eq:Pe_compact}
    P_e(k,n) = \frac{2\Lambda^{k/(n-k)}}{k!}\,
               e^{-\Lambda^{1/(n-k)}}.
\end{equation}
For $k=0$ this gives $P_e(0,n)=2\exp(-(n!)^{1/n})$, which is strictly
decreasing in $n$ since $(n!)^{1/n}\sim n/e\to\infty$, and is maximized at
$n=1$ yielding $2/e$. For $k\geq 1$ one can show that the minimum of
$y=\Lambda^{1/(n-k)}$ (achieved at $n=k+1$, giving $y=k+1$) already
satisfies $P_e(k,k+1)=2(k+1)^k e^{-(k+1)}/k!< 2/e$. Hence $P_e(k,n)\leq P_e(0,1)=2/e$ for all $k<n$, with equality
only at $(k,n)=(0,1)$. The values for the first ten levels are shown in
Fig.~\ref{fig:prob_succ}. It is worth noticing finally that the preprocessing stage generalizes directly to any magic phase. Preparing
$\ket{M_k}=\diag(1,e^{i\pi/(2k)})\ket{+}$ on the Fock elements $\ket{0},\ket{1}$ requires $\alpha=e^{i\pi/(2k)}$,
which maintains $|\alpha|=1$ and hence $P_e=2/e$ for all $n$.

\emph{Stage 2: CNOT gate.}  Once the preprocessing succeeds, we remove the auxiliary qubit from the cavity and then we apply a  CNOT gate where 
the computational atom acts as control (with trigger state $\ket{e}$), and the cavity as a target, i.e. $\mathrm{CNOT}^{(e)}_{a \to c}$.
This is realized by adiabatically transporting the computational atom into the
cavity and applying a  resonant $\pi$-pulse that drives the dressed-state transition
\begin{equation}
 \Pi_2:\quad   \ket{\mathcal{V}_{+}^{(0)}}
 \leftrightarrow\ket{\mathcal{V}_{+}^{(1)}},
\end{equation}
which corresponds in the bare basis to $\ket{e,0}\leftrightarrow\ket{e,1}$: the
cavity photon number is flipped if and only if the atom is in $\ket{e}$.

\emph{Stage 3: Measurement.}  The final stage of the protocol completes the teleportation of the $T$ gate on
the computational atom by measuring the cavity state.
Since the cavity cannot be measured directly, its state must be
transferred to an auxiliary atom (which may be the same atom used in the preprocessing stage upon resetting to $\ket{g}$). The
auxiliary atom is thus transported into the cavity and a new pulse $\Pi_1$ is applied, copying the cavity logical state onto the auxiliary atom. Once the auxiliary atom is extracted from the cavity, its projective measurement is operationally equivalent to measuring the cavity logical state, thereby
completing the teleportation circuit of Fig.~\ref{fig:T-injection} (yellow box in the figure). Depending
on the outcome, either no further operation is required or an $S$ gate is applied to the computational atom. 

\begin{figure}[H]
    \centering
    \includegraphics[width=\columnwidth]{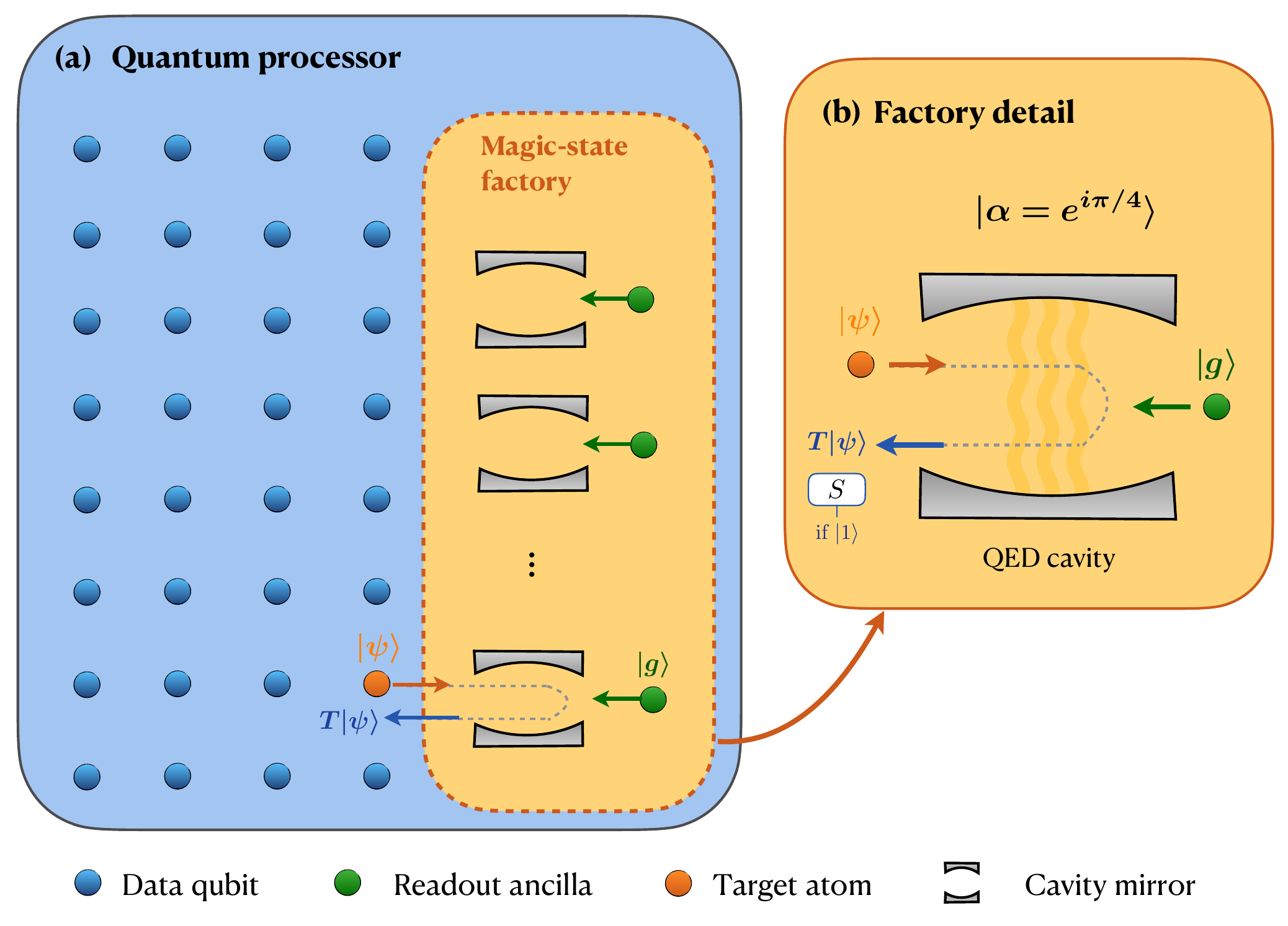}
    \caption{(a) Rydberg atom-based quantum processor incorporating a cavity
    magic-state factory (shaded region; panel (b)) that injects non-Clifford gates
    sequentially or in parallel across the qubit register.}
    \label{fig:architecture}
\end{figure}

\emph{\bfseries Architectural implementation.}
The proposed protocol naturally integrates with state-of-the-art Rydberg-atom quantum computing architectures~\cite{Jaksch2000,Saffman2010,Evered2023,Bluvstein2026,de2026realization,PRXQuantum.3.030314,4c33-b1dv}, 
where all the required ingredients: cavity-mediated interactions, controlled atomic transport, and high-fidelity state detection, have been demonstrated. A single cavity together with a single auxiliary atom suffices to inject $T$ gates sequentially across an entire processor, while parallel operation can be achieved using an array of cavities acting as a magic-state factory, see Fig.~\ref{fig:architecture}. Within such an architecture, a $T$ gate is implemented on an arbitrary target qubit by transporting the corresponding atom into the cavity, where it undergoes the prescribed interaction and is subsequently extracted. An auxiliary atom, reinitialized in $\ket{g}$, is then brought into the cavity to perform the readout CNOT. Following measurement, a conditional $S$ gate completes the operation. 

It is worth stressing that the present protocol operates at
the physical qubit level: each target atom is a single physical
qubit. An important future direction is the extension of cavity-mediated injection to
the \emph{logical} level, where qubits are encoded in a quantum error-correcting
code such as the surface code. One natural route is to use the present
physical-level protocol as a subroutine: the cavity prepares a physical magic
state that is injected through the standard logical injection
protocol~\cite{bravyi2005universal, Litinski2019}. A more direct route, unique to Rydberg platforms, exploits the strong
Rydberg-Rydberg van der Waals interactions and the optical nonlinearities
achievable in atom-cavity systems. In such a scenario, collective Rydberg
excitations across a logical block can mediate the cavity-qubit coupling
required to prepare and inject the magic state at the logical level, without
decomposing the operation into single-physical-qubit gates. Specifically,
multi-qubit Rydberg gates~\cite{Evered2023,Bluvstein2026} already demonstrate
the kind of collective entangling operations that would be needed to couple a
logical qubit to a cavity mode. 

\emph{\bfseries Outlook.}
A near-term experimental priority is the realization of the physical-level
protocol in an existing Rydberg-cavity setup. Augmenting it with error-detection
circuits would allow the effective magic-state fidelity to be characterized and
benchmarked within a fault-tolerant context, setting the stage for the
logical-level generalization outlined above. A longer-term goal is the full
integration of the cavity magic-state factory into a scalable fault-tolerant
architecture, potentially replacing large distillation modules with compact,
cavity-based injection units.

We have proposed and analyzed a cavity-mediated probabilistic magic-state injection protocol. By encoding the magic
state in the $\ket{n=0,1}$ Fock subspace of the cavity, the protocol achieves a
success probability of $2/e\approx 74\%$ per attempt -- independent of the target
magic phase -- with an expected overhead of $e/2\approx 1.36$ attempts per
successful injection. All required operations are available in current
Rydberg atom-cavity platforms. Numerical
simulations confirm the analytical predictions with high accuracy.

The injection is performed at the physical level, and we have further argued that the approach can in principle be generalized to the
logical level: either by using the physical-level protocol as a subroutine
within a fault-tolerant injection and distillation protocol, or by leveraging collective Rydberg
interactions and optical nonlinearities to implement the injection directly on
code-encoded logical qubits. These extensions constitute promising directions
for future theoretical and experimental investigation toward scalable,
distillation-free fault-tolerant quantum computation in neutral-atom systems.

\emph{\bfseries Acknowledgments.}
We acknowledge the use of \texttt{QuTiP}~\cite{qutip5} for numerical simulations.
We acknowledge financial support by MUR (Ministero dell'Universit\`a e della
Ricerca) through the PNRR MUR project PE0000023-NQSTI.

\bibliography{biblio}

\onecolumngrid
\appendix

\section{Simulation parameters and numerical validation}
\label{app:simulation}

\begin{figure}[H]
    \centering
    \includegraphics[width=0.5\linewidth]{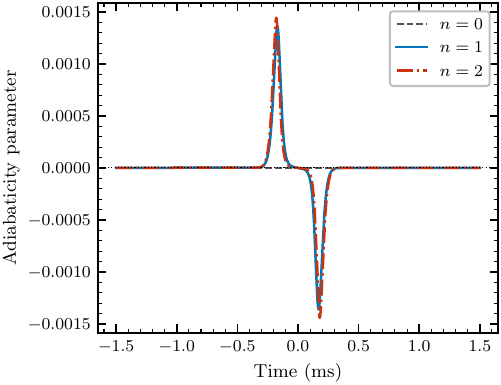}
    \caption{Left-hand side of the adiabatic condition
    Eq.~\eqref{equation:adiabatic_condition} for dressed states $n=0$ (dashed), $n=1$ (solid blue), and $n=2$ (dash-dot red). The condition is satisfied to within three orders of magnitude.}
    \label{fig:adiabatic_condition_app}
\end{figure}

The simulation follows the CNOT implementation of Ref.~\cite{Giovannetti2000}
with parameters:
\begin{equation}
    \Omega_0\sim 420\,\mathrm{kHz},\quad
    \tau\sim 100\,\mu\mathrm{s},\quad
    \delta\sim 0.18\,\Omega_0,
\end{equation}
satisfying the adiabatic condition
\begin{equation}
    \label{equation:adiabatic_condition}
    \frac{\dot{\Omega}_{\rm{eff}}(t)\,\delta\sqrt{n}}
         {4\bigl[(\delta/2)^2+\Omega_{\rm{eff}}^2(t)\,n\bigr]^{3/2}}
    \ll 1\;,
\end{equation}
to within three orders of magnitude throughout the protocol, as shown in
Fig.~\ref{fig:adiabatic_condition_app} for the first three dressed states.

The two $\pi$-pulses address dressed-state transitions with different
energy splittings $E_+^{(1)}-E_-^{(0)}$ and $E_+^{(0)}-\hbar\omega_0$
at the operating point $(\Omega_0,\delta)$. Because the driving field
Eq.~\eqref{equation:CNOT_pulse} couples to the dressed states through their
bare-state decomposition, the effective Rabi frequency for each transition is
proportional to the product of the corresponding dressed-state amplitudes,
which depends on the mixing angle at the operating point. At
$\delta\sim 0.18\,\Omega_0$ this yields a ratio
$\Theta_0/\Theta_1\approx 0.914$, computed numerically from the
dressed-state vectors of Eq.~\eqref{eq:dressed}. The CNOT pulses [Eq.~\eqref{equation:CNOT_pulse}] use the parameters
\begin{align*}
 &\Theta_0\sim 129.33\,\mathrm{kHz},\quad
     \tau_0\sim 3.8\,\mu\mathrm{s},\\
    &\Theta_1\sim 141.5\,\mathrm{kHz},\quad
     \tau_1\sim 19\,\mu\mathrm{s},
   \end{align*}
where the subscript denotes the control Fock level. The CNOT controlled
by $\ket{1}$ is activated at $t=0$; the CNOT controlled by $\ket{0}$ follows
$\sim 30\,\mu\mathrm{s}$ later.

\end{document}